\documentstyle[12pt]{article}
\newcounter{multieqs}
\newcommand{\eqnumber}{\addtocounter{multieqs}{1}
\addtocounter{equation}{-1}}
\newcommand{\begalph}
{\setcounter{multieqs}{0}
\addtocounter{equation}{1}
\renewcommand{\theequation}
{\arabic{equation}.\alph{multieqs}}}
\newcommand{\alphend}
{\setcounter{multieqs}{0}
\renewcommand{\theequation}{\arabic{equation}}}

\newcommand{\N}{{\rm I}\hspace{-0.19 em}{\rm N}}
\begin{document}

\begin{center}
\Large{CONTINUOUS FAMILY OF EINSTEIN-YANG-MILLS WORMHOLES}
\end{center}
\begin{center}
        {\bf E.E. Donets}\\[0.2cm]
	 Department of Theoretical Physics, Physics Faculty,\\
         Moscow State University, 119899 Moscow, Russia\\[0.5cm]
	{\bf D.V. Gal'tsov}\footnote{On leave from the
	 Department of Theoretical Physical, Moscow State University and
	 from the University of Campinas, Brazil.} \\[0.2cm]
	 	 Instituto de F\'{\i}sica, Universidade de S\~ao Paulo, \\
	 CP 20.516, 01498-970, S\~ao Paulo, SP, Brasil
\end{center}

\vspace{0.5cm}

\begin{center}
{\bf Abstract}
\end{center}

     It is shown that for some particular value of the cosmological
constant depending on the gauge coupling constant a continuous
one-parameter family of Einstein-Yang-Mills wormholes exists which
interpolates between the instanton and the gravitating meron
solutions. In contradistinction with the previously known solutions the
topological charge of these wormholes is not quantized. For all of
them the contribution of gravity to the action exactly cancels that
of the gauge field.

\vfill \eject

     Euclidean space-time wormholes with Yang-Mills (YM) fields
\cite{r1}--\cite{r6} are obtained as solutions of the coupled
Einstein-Yang-Mills
(EYM) system of equations with non self-dual YM fields. In the flat
euclidean space pure gauge fields can form structures known as an
instanton, a meron, and nested, or dressed, merons interpolating
between the first two \cite{r7}. They can be labeled by the topological
charge $Q$ that varies continuously between the value $1/2$ for the meron
and $1$ for the instanton. The instanton is self-dual and has a zero
energy-momentum tensor, hence the gravity can be added in a
self-consistent way just as a background field. The energy-momentum
tensor of the meron is non-zero and this changes the situation
dramatically. When gravity is coupled, the singularity at the
location of a meron expands to a wormhole throat, and, consequently,
the euclidean topology of the space-time transforms to that of a
wormhole \cite{r1}. The value of the topological charge of the meron
wormhole turns out to be zero, the charge of the meron being
``swallowed'' by the wormhole \cite{r3}.

      Unfortunately, the total action of this solution is divergent
because of the slow fall-off of the meron field at infinity. To
remedy this in a physically motivated way, a positive cosmological
constant can be added. Then the asymptotic behavior of the field
will be irrelevant since the corresponding action becomes finite simply
due to the compactness of the space. Such solutions could be
interpreted as describing the tunneling from the hot Friedmann-
Robertson-Walker (FRW) Universe to the De Sitter one \cite{r6}.

      Now, the addition of the cosmological constant introduces one
more new feature comparatively with the flat space case. The family
of solutions with the topological charge between that of the
gravitating meron $(Q=0)$ and the instanton on the FRW background $(Q=1)$
was described by Verbin and Davidson
\cite{r2} and Rey \cite{r4}. These solutions
could pretend to be nested meron wormholes. However, in contradistinction
with the flat-space nested merons, these gravitating solutions
can posses only some discrete values of the topological charge. This
comes about as follows. For the isotropic homogeneous SU(2) (YM)
field and the corresponding FRW metric the Einstein-Yang-Mills (EYM)
system of equations reduces to a two-dimensional separable non-linear
oscillator system. In order to be interpreted as a wormhole, the
solution must be strictly periodic, i.e. the periods of two oscillators
have to be in some rational relation. This imposes a quantization
condition on the separation constant, which can be related to
the topological charge. So the proposed nested meron wormholes have a
quantized topological charge and consequently do not interpolate
continuously between the (self-dual) instanton and the meron
solutions as nested merons do in the flat space.

Here we will show, however, that some critical value of the cosmological
constant depending on the gauge coupling constant exists
for which both oscillators have equal periods for all values of the
separation constant between those of the meron and the instanton.
These wormholes can have all values of the topological charge on the
interval (0, 1), and hence form indeed a continuous family of
solutions interpolating between the instanton and the meron. The
striking feature of these solutions is that the corresponding total
action is exactly zero, the positive contribution of the YM
field being canceled by the negative contribution of the
gravitational field.

      The critical value of the cosmological constant turns out to be
the same as was found by De Alfaro, Fubini and Furlan \cite{r8} in the
context of a strong gravity theory and subsequently was used in the
induced gravity approach \cite{r9}--\cite{r10}.
We show that following the lines of
\cite{r8}--\cite{r9} a particular solution belonging to our complete set of
solutions can be obtained also. However it does not have the
desired wormhole properties because of inappropriate
boundary conditions intrinsic to this approach.

     We start with the action for the Euclidean EYM system with the
cosmological constant
\vspace{0.3cm}
\begin{equation}
S = \frac{1}{16\pi} \, \int \, \sqrt{g}\, (m^2_{p\ell}(2\lambda - R) +
F^a_{\mu\nu} \, F^{a{\mu\nu}}) \, d^4x \quad ,
\end{equation} 

\vspace{0.3cm}

\noindent assuming the metric to have the form

\begin{equation}
ds^2 = N^2 \, d\tau^2 + a^2 (d\chi^2 + \sin^2\chi \, (d\theta^2 +
\sin^2\theta d\varphi^2)) \quad .
\end{equation} 

     For the SU(2) YM connection (generalization to higher groups is
straightforward along the lines of \cite{r5}--\cite{r6}) symmetric under the
space-time isometry group SO(4) one can use either a temporary gauge
\cite{r1}--\cite{r2} or a Witten ansatz,
the relationship between the two having
been clarified in \cite{r11}. The field parametrized by a single
real-valued function $s$ of the euclidean time $\tau$ in the temporary
gauge reads
\begin{eqnarray}
A_0 & = & 0 \quad , \qquad  eA_j = \frac{i}{2} \, (\sigma(\tau) + 1) \,
U{\partial_j}U^{-1} \quad ; \nonumber  \\
U & = & \exp \, [i\chi \, (\sin\theta(\sigma_x\cos\varphi +
\sigma_y\sin\varphi)
 + \sigma_z\cos\theta)] \quad .
\end{eqnarray} 

     Integrating out the angle variables in (1) and  eliminating  the
total derivative one gets $S = S_{gr} + S_{YM}$,
\begalph
\begin{eqnarray}
S_{YM} & = & \frac{3\pi}{4e^2} \, \int \, \left(\frac {\dot{\sigma}^2a}{N}
+ (\sigma^2 - 1)^2 \, \frac{N}{a} \right) \, d\tau \quad , \eqnumber \\
S_{gr} & = & \frac{3\pi m^2_{p\ell}}{4} \, \int \, \left(
\frac{-{\dot{a}^2a}}{N} - aN + \frac{\lambda}{3} \, Na^3 \right) \,
d{\tau} \quad , \eqnumber
\end{eqnarray} 
\alphend
\noindent where dot denotes the derivative with respect to $\tau$.
The corresponding equations of motion for the YM variable
$\sigma(\tau)$ and the radius $a(\tau)$ of the Universe read
\begalph
\begin{eqnarray}
& & \frac{d}{d\tau} \,\left(\frac{\dot\sigma a}{N} \right) -
2{\sigma} \, (\sigma^2 -1) \, \frac{N}{a} = 0 \quad , \eqnumber \\
& & \frac{d}{d\tau} \, \left(\frac{\dot{a}a}{N}\right) - N +
\frac{2}{3} \, \lambda Na^2 = 0 \quad . \eqnumber
\end{eqnarray} 
\alphend
\noindent Variation with respect to the lapse function gives the constraint
\vspace{0.3cm}
\begin{equation}
m^2_{p\ell} \, \left(\frac{a{\dot a}^2}{N^2} - a + \frac{2}{3} \,
\lambda a^3 \right) + \frac{1}{e^2} \, \left(- \frac{\dot{\sigma}^2 a}{N^2} +
\frac{(\sigma^2 - 1)^2}{a} \right) = 0
\end{equation} 

\vspace{0.3cm}

\noindent which is just the $\tau\tau$ - component of the Einstein
equations. The non-linear two-oscillator system (5) separates in the
conformal gauge
$N=a$:
\begalph
\begin{eqnarray}
& & \ddot{\sigma} - 2\sigma \, (\sigma^2 -1) = 0 \eqnumber \\
& & \ddot{a} - a + \frac{2}{3} \, \lambda a^3 = 0 \eqnumber
\end{eqnarray} 
\alphend
\noindent while the constraint equation (6) then just equates the first
integrals of (7)
\begalph
\begin{eqnarray}
& & \dot{\sigma}^2 + U_{\sigma} = -C \quad , \qquad U_{\sigma} = - \,
(\sigma^2 -1)^2 \quad , \eqnumber \\
& & \dot{a}^2 + U_a = - \frac{C}{m^2_{p\ell}e^2} \quad , \qquad
U_a = \frac{\lambda a^4}{3} - a^2 \quad .\eqnumber
\end{eqnarray} 
\alphend
\noindent Here $C$ is the separation constant, restricted to the interval
$[0,1]$ for the wormhole solutions \cite{r2}.

     The quatric potentials $U_{\sigma}$ and $U_{a}$ have a similar shape and
possess four zeros
\begalph
\begin{eqnarray}
& & U_{\sigma} \, (\pm\sigma_{\pm}) = 0 \quad , \qquad
\sigma_{\pm} = (1 \pm C^{1/2})^{1/2} \quad , \eqnumber \\
& & U_a \, (\pm a_{\pm}) = 0 \quad , \qquad a_{\pm} =
\left[\frac{3}{2\lambda} \, \left(1 \pm  \left(1 - \frac{4\lambda
C}{3m^2_{p\ell}e^2}\right)^{1/2}\right) \right]^{1/2} \,\, , \eqnumber
\end{eqnarray} 
\alphend
\noindent satisfying the following relations
\begalph
\begin{eqnarray}
& & \sigma^2_+ + \sigma^2_- = 2 \quad , \qquad \sigma_+
\sigma_- = (1 - C)^{1/2} \quad , \eqnumber \\
& & a^2_+ + a^2_- = \frac{3}{\lambda} \quad , \qquad
a^2_+ a^2_- = \frac{3C}{\lambda m^2_{p\ell}e^2}
 \quad . \eqnumber
\end{eqnarray} 
\alphend

The actual motion for wormhole solutions is restricted to the
intervals $-\sigma_-<\sigma<\sigma_-,\,a_-<a<a_+$, and the boundary conditions
have to be of the following form
\begalph
\begin{eqnarray}
& & \sigma (0) = -\sigma_- \; \;  , \quad \dot{\sigma} (0) = 0 \; \;  ;
\quad \sigma (\tau_f) = \sigma_- \; \;  , \quad
\dot{\sigma} (\tau_f) = 0 \quad ; \eqnumber \\
& & a (0) = a_- \; \; , \quad \dot{a} (0) = 0 \; \; ; \quad
a(\tau_f) = a_+ \; \;  , \quad \dot{a}(\tau_f) = 0  \eqnumber
\end{eqnarray} 
\alphend
\noindent for some moment $\tau_{f}$ of the conformal time $\tau$.
This is, of course, an
overdetermined system, which can not be generally satisfied, thus
leading to the above quantization condition. Indeed,
general solutions of the equations (7), (8) satisfying the initial
conditions (11a) at the left turning points can be expressed in terms
of the elliptic functions as follows
\begalph
\begin{eqnarray}
& &\sigma(\tau) = \sigma_- {\rm sn} \left[\sigma_+ \, \left(\tau -
\frac{T_\sigma}{2}\right) \, ,
k^2_{\sigma} \right] \quad , \qquad
k_{\sigma} = \frac{\sigma_-}{\sigma_+}  \quad ; \eqnumber \\
& & a^2(\tau) = a^2_+ + (a^2_-  - a^2_+) \,{\rm sn}^2
\left[\frac{a_+}{\sqrt{a^2_+ + a^2_-}}\,\left(\tau -T_a \right) \, , k^2_a
\right] \quad,\nonumber \\
& & k_a = \frac{\sqrt{a^2_+ - a^2_-}}{a_+} \quad .\eqnumber
\end{eqnarray} 
\alphend
\noindent The conditions (11b) at the right turning points, however,
can only be satisfied if $\tau_{f} = n_{\sigma}T_{\sigma}$
and $\tau_{f} = n_{a}T_a$ where $T_{\sigma}$ and $T_{a}$ are
proper half-periods for the corresponding variables, and $n_{\sigma},
n_{a}\in\N$.
The periods can be expressed in terms of the complete elliptic integrals
of the first kind
\begalph
\begin{eqnarray}
T_{\sigma} & = & \frac{2}{\sigma_+} \,  K \, (k_{\sigma}) \quad , \eqnumber \\
T_a & = & \frac{\sqrt{a^2_+ + a^2_-}}{a_+} \, K \, (k_a) \quad , \eqnumber
\end{eqnarray} 
\alphend
\noindent and we thus get the quantization condition  of  Verbin  and
Davidson \cite{r2} for the separation constant $C$
\begin{equation}
n_{\sigma} \, T_{\sigma} = n_a \, T_a \quad .
\end{equation} 

      Let us show that the Eq.(14) can be thought of as a quantization
condition for the topological charge
\begin{equation}
Q = \frac{e^2}{64\pi^2} \, \int^{T_\sigma}_{0} \, d\tau \,
\int_{S^3} \, e^{\mu\nu\lambda\tau} \, F^a_{\mu\nu} \, F^a_{\lambda\tau} \,
d^3x = P(T_{\sigma}) - P(0) \quad ,
\end{equation} 
\noindent where
\begin{equation}
P(\tau) = \frac{e^2}{32\pi^2} \, \int_{S^3} \, e^{o\mu\nu\lambda} \, A^a_{\mu}
\left(\partial_{\nu}A^a_{\lambda} + \frac{e}{3} \, e^{abc} \, A^b_{\nu}
\, A^c_{\lambda} \right) \, d^3 x
\end{equation} 
is the Chern-Simons number. Substituting the Eq.(3) we get
\begin{equation}
P(\tau) = \frac{1}{4} \, (3\sigma + 2 -{\sigma}^3) \quad ,
\end{equation} 
and, hence,
\begin{equation}
Q = \frac{1}{2} \, \sigma_- \, (3 - \sigma^2_-) \quad .
\end{equation} 
\noindent Because of the quantization of $C$, this quantity will have also
discrete values.

      Now, if we  take  into  account  the  following  transformation
formula for the complete elliptic integrals of the first kind
\begin{equation}
K  (2\sqrt{z}/(1+z)) = (1 + z) \, K \, (z)
\end{equation} 

\noindent we can easily realise that the necessary relation between the moduli
$k_{\sigma}$ and $k_a$ of the elliptic functions,
entering the formulas (13) can
be satisfied indeed if only the turning points are related by
\begin{equation}
m_{p\ell} \, e \, a_{\pm} = \sigma_+ \; {\pm} \; \sigma_-\quad .
\end{equation} 

\noindent This fix the value of the cosmological constant as follows
\begin{equation}
\lambda = \lambda_{cr} = \frac{3}{4} \, m^2_{p\ell} \, e^2 \quad .
\end{equation} 

\noindent For this particular value of the cosmological constant the relation
\begin{equation}
k_a = \frac{2\sqrt{k_\sigma}}{1 + k_{\sigma}}
\end{equation} 
\noindent holds for all $C\in [0,1]$, and the half-periods exactly coincide
$T_{\sigma} = T_a = T$. It is interesting to note that for
such critical $\lambda$ the
radius $a$ becomes constant for $C=1$, i.e. in the meron limit. Hence in
our case the meron wormhole metric transforms into that of the
Euclidean static Einstein Universe. For $C $ non equal to $ 1,0$  the solution
(12)
preserves its meaning of a wormhole.

     Let us calculate the value of the action integral between  $\tau = 0$
and $\tau = T$. Changing the integration variable
in the Eq.(4a) to $\sigma$  one
can easily obtain the desired  quantity  in  terms  of  the  complete
elliptic integrals of the first and the second kind
\vspace{0.3cm}
\begin{eqnarray}
S_{Y M} & = & \frac{3\pi}{4e^2} \, \int^T_0 \,
(\dot{\sigma}^2 - U_{\sigma}) \, d\tau \nonumber \\
& = & \frac{2\pi}{e^2} \, \left(\sigma_+ E(k_{\sigma}) -  \left(C +
\frac{\sqrt{C}}{4} \right) \, \frac{1}{\sigma_+} \, K (k_{\sigma})
\right) \quad .
\end{eqnarray} 

\vspace{0.3cm}

\noindent Similarly, the gravitational part of the action is calculated passing
to the integration variable $a$
\vspace{0.3cm}
\begin{eqnarray}
S_{g\tau} & = & \frac{3\pi m^2_{p\ell}}{4} \, \int^T_0 \, (- {\dot a^2} +
U_a) \, d\tau \nonumber \\
& = & \sqrt\frac{3}{\lambda}  \, \frac{\pi m^2_{p\ell}}{2} \,
\left( - a_+ \, E \, (k_a) + \frac{1}{2m^2_{p\ell}e^2a_+} \, C \, K \,
(k_a) \right) \quad .
\end{eqnarray} 

\vspace{0.3cm}

\noindent Now, for the critical cosmological constant (21)  the  relation (22)
between the arguments of elliptic integrals holds. To relate the
values (23) and (24) one has to use the transformation formula (19)
together with the corresponding formula for the elliptic integrals of
the second kind $(0\leq z\leq 1)$:
\begin{equation}
(1 + z) \, E \, \left(\frac{2\sqrt{z}}{1 + z} \right)
= 2 \, E(z) - (1 - z^2) \, K(z) \quad .
\end{equation} 

\noindent This gives precisely
\begin{equation}
S_{Y M} + S_{gr} = 0 \quad .
\end{equation} 

To get more physical insight into the nature of the critical $\lambda$
given by the Eq.(21) we turn to the  discussion  of  the  conformal
properties of the solutions in question. Since the YM  equations  are
conformally invariant, and we are considering  conformally  flat  (in
fact conformal to the static Einstein Universe)  metrics,  one  could
start with the YM theory in the flat Euclidean  space-time.  Then the
standard Corrigan-Fairly-'t Hooft-Wilczek ansatz (in cartesian
coordinates $t, x, y, z$)
\begin{equation}
A_{\mu} = \frac{i}{e} \, \sigma_{\mu\nu} \, \partial^{\nu} \, \ln \, h \quad ,
\end{equation} 

\noindent where ${\displaystyle \sigma_{oi} =
\frac{\sigma_i}{2}, \sigma_{ij} =
\frac{1}{2}\epsilon_{ijk} \, \sigma^k}$ \,, provides a solution of the
sourceless YM
equation if the following equation for the scalar function $h$ holds
\begin{equation}
\Box h + \kappa \, h^3 = 0 \quad ,
\end{equation} 

\noindent where $\kappa$ is  some  (still arbitrary) constant, and $\Box$ is a
flat-space four-dimensional Laplacian. Now, as De Alfaro, Fubini and
Furlan had found \cite{r8} in the context of the strong gravity, the theory
may be put into a larger curved space context by considering
confomally flat metrics
$g_{\mu\nu} = h^2 \, \delta_{\mu\nu}$ with the same function
entering as a conformal factor. The corresponding scalar curvature will
be
\begin{equation}
R = - \frac{6\Box h}{h^3} \quad .
\end{equation} 

\noindent Then taking into account the
tracelessness of the  YM  energy-momentum
tensor, one gets from the Einstein equations  with  the  cosmological
constant $\lambda$ the following equation for $h$:
\begin{equation}
\Box h + \frac{2}{3} \, \lambda h^3 = 0
\end{equation} 

\noindent This is just the equation (28) with a particular value of the
constant $\kappa = 2\lambda/3$. Now, from
the gravitational Hamiltonian constraint
equation we get exactly our relation between the cosmological
constant and the gauge coupling constant (21).

      To relate this to our previous considerations, let $h$  be a
function of the variable $\tau = (t^2 + x^2 + y^2 + z^2)^{1/2}$ and make the
following coordinate transformation
\begin{eqnarray}
\begin{array}{lll}
t & = & \cos\chi\exp(\tau) \quad , \\
x & = & \sin\chi\cos\varphi\sin\theta\exp(\tau) \quad , \\
y & = & \sin\chi\sin\varphi\sin\theta\exp(\tau) \quad , \\
z & = & \sin\chi\cos\theta\exp(\tau) \quad .
\end{array}
\end{eqnarray} 

\noindent This will reproduce our previous ansatze
for  the  metric  and  the  YM connection (2), (3) with
\begin{equation}
\sigma = - 1 -\dot{h}/h \quad , \qquad  a = h\exp(\tau)
\end{equation} 

\noindent where a dot denotes the derivative with respect to $\tau$
as before.  From
here it follows immediately that
\begin{equation}
\sigma = - \frac{\dot{a}}{a}
\end{equation} 

      Since both the conformal factor and the YM connection were  now
described in terms of the only function $h$, it is clear that both
variables will have the same periods for periodic solutions. Hence
we have reproduced the  desired  property  imposed  by  the  wormhole
boundary  conditions  (11)  from an  alternative  point  of  view.
However, the equation (33) selects from the whole set of solutions of
the system (7), (8) a particular one, which can not be interpreted
itself as a wormhole, because the initial conditions according to the
Eq (29) will be $\sigma(0) = - {\dot a}(0)/a(0)$ (i.e. when thy conformal
factor starts tunneling, the YM field starts exactly in the middle
under the potential barrier). Yet this particular solution seems to
be interesting by itself. Amazingly enough it can be described
entirely in terms of elementary functions. Let us use comoving
coordinates in (2) putting in the Eqs(5) $N=1$. Making a substitution
$f = a^2$ we will get a linear equation
\begin{equation}
\frac{d^2f}{d\tau^2_c} - 2 + (e m_{p\ell})^2 \, f = 0
\end{equation} 

\noindent which is solved with initial conditions
$f(0) = a^2_-,{\dot f}(0) = 0$ as follows

\begin{equation}
f = \frac{a^2_+ + a^2_-}{2} - \frac{a^2_+ - a^2_-}{2}
\cos(e m_{p\ell}\tau_c) \quad .
\end{equation} 

\noindent The corresponding YM function can then be obtained from Eq.(29):
\begin{equation}
\sigma = -  \frac{1}{2\sqrt{f}} \, \frac{df}{d\tau_{c}} \quad .
\end{equation} 

      To summarize: we have shown, that a continuous family of EYM
wormholes with the gauge group SU(2) exists for some particular
value of the cosmological constant depending on the gauge coupling
constant. The absolute value of the total action in this case is
minimal and moreover, exactly equal to zero. The meron wormhole then
transforms into the static Einstein Universe, while nested merons
correspond to wormholes possessing all values of a topological charge
between $0$ and $1$. Cosmological implications of the results obtained
will be discussed in a separate publication.

D.V.G. thanks the Institute of Physics the University of S\~ao Paulo (Brazil)
for the hospitality.

\pagebreak


\begin{thebibliography}{99}
\bibitem{r1} A. Hosoya and W. Ogura, Phys. Lett., {\bf 225B} (1989) 117.
\bibitem{r2} Y. Verbin, A. Davidson, Phys. Lett.,
{\bf 229B} (1989)  364;  Nucl.  Phys. 339B (1990) 545.
\bibitem{r3} A.K. Gupta, J. Hughes, J. Preskill, M.B. Wise,
Nucl.  Phys. {\bf 333B} (1990) 195.
\bibitem{r4} S.J. Rey, Nucl. Phys. {\bf 336B} (1990) 146.
\bibitem{r5} K. Yosida, S. Hirenzaki, K. Shiraishi,
Phys. Rev. {\bf D42} (1990) 1973.
\bibitem{r6} O. Bertolami, J.M.  Mourao,  R.F. Picken,
I.P. Volobujev,  Int.  J. Mod. Phys. {\bf A6} (1991) 4149.
\bibitem{r7} C.G.Callan, R. Dashen, D.J. Gross,
Phys. Rev. {\bf D17} (1978) 2717.
\bibitem{r8} V. De Alfaro, S. Fubini, S. Furlan,
Nuovo Cim. {\bf A50} (1980) 523.
\bibitem{r9} J. Cervero , Gen. Rel. and Grav.
{\bf 14} (1982) 393; Phys. Lett. {\bf 108B} (1982) 108.
\bibitem{r10} J. Cervero, P.G. Estevez, Ann. Phys. {\bf 142} (1982) 64.
\bibitem{r11} D.V. Gal'tsov, M.S. Volkov, Phys. Lett.  {\bf 256B} (1991) 17.
\end{thebibliography}
\end{document}